\documentclass{article}

\usepackage{arxiv}
\usepackage{float}

\usepackage[utf8]{inputenc} 
\usepackage[T1]{fontenc}    
\usepackage{graphicx}       
\usepackage{hyperref}       
\usepackage{url}            
\usepackage{microtype}      
\usepackage[numbers,sort&compress]{natbib} 
\usepackage{booktabs}
\usepackage{tabularx}
\usepackage[table]{xcolor}   
\usepackage{tikz}            

\newcommand{\hbr}{0.62ex}    
\newcommand{\ballfull}{\tikz[baseline=-0.6ex]{\fill (0,0) circle (\hbr);}}
\newcommand{\ballempty}{\tikz[baseline=-0.6ex]{\draw[line width=0.4pt] (0,0) circle (\hbr);}}
\newcommand{\ballhalf}{\tikz[baseline=-0.6ex]{%
  \draw[line width=0.4pt] (0,0) circle (\hbr);
  \begin{scope}\clip (-\hbr,-\hbr) rectangle (0,\hbr);\fill (0,0) circle (\hbr);\end{scope}}}

\title{Who Pays When Shared Infrastructure Fails? \\[0.45em]
  \large Zero Liquid Discharge, the Utilisation Trap, and the Incidence of
  Compliance Cost in India's Textile and Tannery Clusters}

\author{ \href{https://orcid.org/0009-0002-7864-4642}{\includegraphics[scale=0.06]{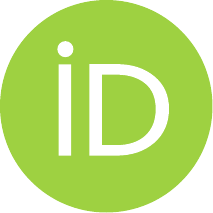}\hspace{1mm}Mihika Singhania}}

\renewcommand{\shorttitle}{Who Pays When Shared Infrastructure Fails?}

\hypersetup{
pdftitle={Who Pays When Shared Infrastructure Fails?},
pdfsubject={Environmental regulation, water pollution, industrial clusters},
pdfauthor={Mihika Singhania},
pdfkeywords={Zero liquid discharge, Common effluent treatment plants, Regulatory enforcement},
}

\begin{document}
\maketitle

\begin{abstract}
Many countries require factories to recycle almost all their wastewater and return none to rivers, a standard called zero liquid discharge (ZLD); in India, courts have imposed it on textile and leather clusters. Yet when enforcement tightens, small firms close while larger polluters often endure. This paper asks why, comparing Tirupur, where closures have already occurred, with Kanpur's Jajmau tanneries, where they are ongoing. Small firms depend on shared treatment plants they do not control. As these plants become underused, costs rise, performance declines, and firms with in-house treatment are better able to withstand enforcement. Because compliance is judged by equipment ownership rather than measured discharge, plant failure is attributed to member firms. This final step is the best evidenced, while the earlier links remain suggestive. The paper recommends judging compliance by measured discharge and determining whether the plant or the firm has failed before closure, so enforcement targets pollution rather than the firms least able to absorb it.
\end{abstract}


\section{Introduction}
\label{sec:intro}

In early 2011, the Madras High Court ordered the closure of Tirupur's dyeing and bleaching units, together with the common effluent treatment plants that served them, ruling on a contempt petition brought by farmers downstream on the Noyyal River \citep{gronwall2017impact}. Around 700 units stopped work. Estimates of the resulting job losses reached 300,000. The number dwarfs the direct workforce of the closed units, because dyeing and bleaching are the processing stage the whole knitwear cluster depends on, so their closure idled ancillary and downstream work across the town. That number comes from contemporary press accounts rather than any official survey \citep{gronwall2017impact}. Reopening was slow, taking up to two years for most units and longer for several of the plants, and close to 100 units never returned. Some production shifted to dyeing houses in Ludhiana and Surat, and some of it, according to local accounts, resumed quietly at micro-scale inside residential buildings that discharged into the municipal sewer. The order had been meant to keep untreated effluent out of the river, yet part of what it produced was effluent that was harder to trace and harder to treat.

Those units were closed for failing to meet zero liquid discharge (ZLD), the requirement that a plant recover almost all of its process water and return none of it to surface water. It is, in effect, a circular-economy standard for water, closing the loop inside the factory rather than discharging to the river. The standard responds to a real harm, since the wet processing of cloth and hides releases saline, chemically loaded effluent that has degraded Indian rivers for decades. In the textile clusters it arrived less through a single national statute than through litigation, with the Madras High Court prescribing its technical form for Tirupur in the mid-2000s. Compliance is expensive, and the expense is lumpy rather than smoothly scalable: the evaporators and crystallisers needed to capture the last portion of the liquid stream have been estimated to roughly double the cost of a system that stops at reverse osmosis. Such systems remain characteristic of larger units running their own plants \citep{gronwall2017impact}. Faced with this indivisibility, the regulatory system chose to share it. The Central Pollution Control Board's own account of why common effluent treatment plants exist is plain, namely that they serve clusters of small units because such units cannot bear the capital and maintenance costs alone \citep{cpcb2014tirupur}. Small firms did not choose shared treatment; the system built it for the firms it had priced out of treating their own effluent.

\begin{figure}[H]
    \centering
    \includegraphics[width=0.85\linewidth]{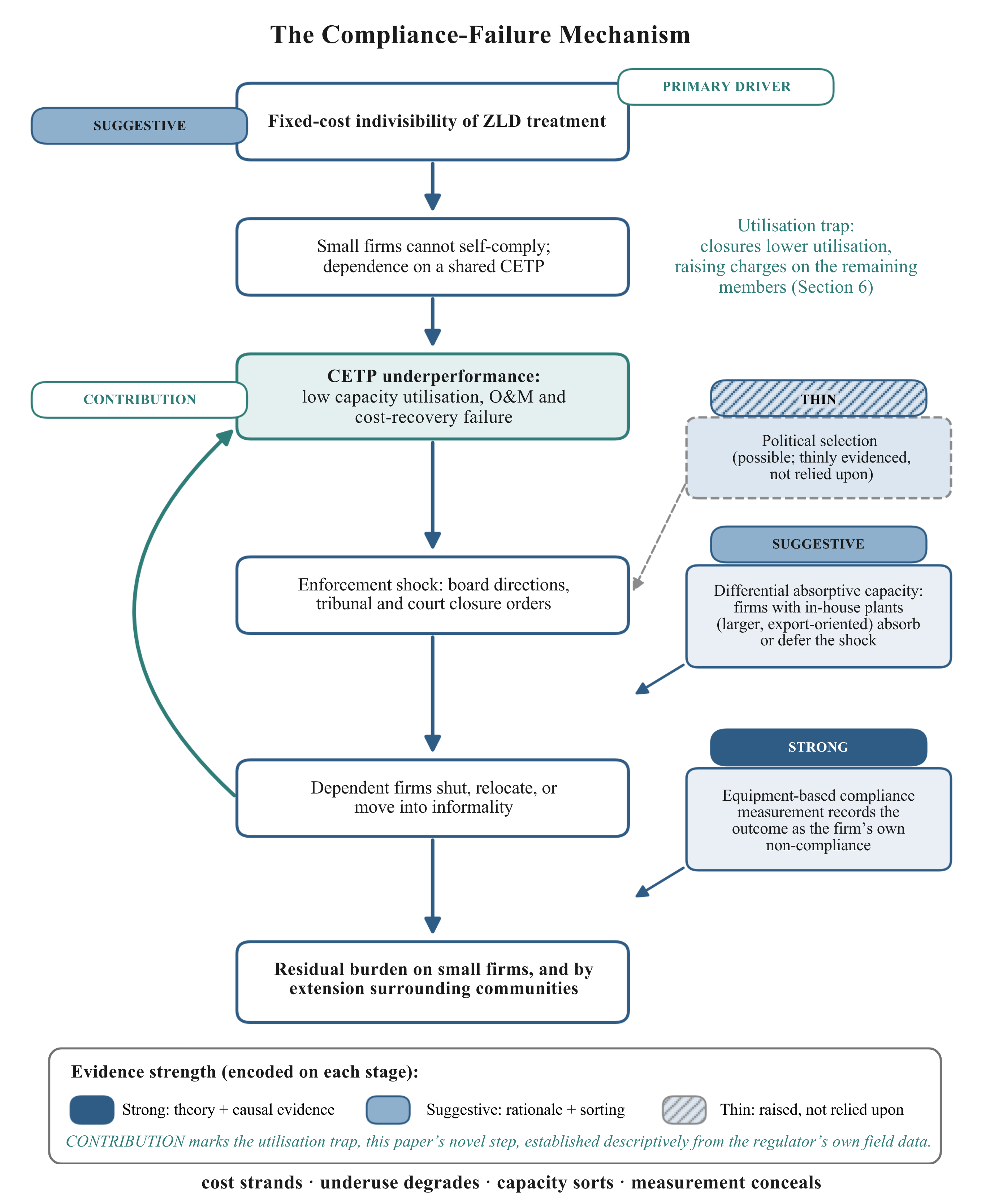}
    \caption{\textit{The compliance-failure mechanism.} Cost indivisibility strands small firms on a shared plant they cannot control; low utilisation degrades that plant; differential absorptive capacity sorts which firms survive an enforcement shock; and equipment-based measurement records the resulting failure as the firm's own. The left-hand feedback arrow marks the utilisation trap, and each stage is tagged with the strength of its supporting evidence.}
    \label{fig:mechanism}
\end{figure}
That arrangement carries a liability which has drawn little scrutiny. A firm connected to a shared plant is obliged to use infrastructure it neither owns nor can make perform, while its compliance is judged one unit at a time, so that a failure of the plant is recorded as a failure of the firm. The remedy devised for small-firm disadvantage becomes the channel through which small firms are exposed. This goes beyond the familiar point that a fixed-cost mandate weighs more heavily on small producers, which anyone would predict. The harder question is why the liability was placed on the party with no control over the outcome, and the answer lies partly in the institutions responsible, since a court adjudicating river pollution is not equipped to weigh which firms would end up bearing the financial cost of compliance. The paper therefore asks why, when ZLD enforcement tightens in India's textile and tannery clusters, the burden of adjustment falls disproportionately on small firms, and by what mechanism it does so.

The account developed here treats the outcome as one connected process rather than a set of separate causes (Figure~\ref{fig:mechanism}). Because the treatment train cannot be shrunk in proportion to output, small units are drawn into shared plants. The economics of those plants then hinge on how fully they are used. In the Tirupur data, the charge per kilolitre varied with utilisation rather than plant size, the two most underused plants being much the dearest \citep{cpcb2014tirupur}. A trap follows from this, since members who default or leave push utilisation lower, which raises charges on those who stay and drives the plant beneath the throughput it needs to meet its own standards. When enforcement arrives, firms large enough to have built their own plants, usually the export-oriented ones, can absorb the shock where dependent firms cannot. Where compliance is finally recorded as the possession of working equipment rather than as measured discharge, the closures enter the record as the small firm's own fault. Cost strands the small firm, underuse degrades the shared plant, capacity sorts who survives the shock, and measurement records the failure as the firm's own.

These stages do not rest on equal evidence, and are weighted accordingly. The measurement stage is the best supported, resting on established theory and experimental evidence from India \citep{besanko1987performance,duflo2013truthtelling}. The cost stage is weaker, since no measured cost curve against plant capacity exists and it leans on the regulator's rationale and the observed sorting of firms \citep{cpcb2014tirupur,goldar2001water}. Weakest of all is the possibility that political connection independently decides who is spared, which the paper raises but does not lean on \citep{duflo2018value}. Section~\ref{sec:mechanism} takes each in turn.

The question is not only historical. The same configuration is now shaping enforcement in Kanpur's Jajmau tannery cluster, where the shared plant has repeatedly been recorded as non-complying \citep{cpcb2017annual,cag2017report} and dependent units sit under active tribunal and board pressure to close \citep{ngt2021jajmau}. Those decisions are being taken now, on the very compliance record the mechanism identifies as misleading, a case Section~\ref{sec:mechanism} returns to.

Three contributions follow. The paper recasts the observation that pollution-control policy is not automatically fair as a traceable institutional mechanism; it moves the unit of analysis from community exposure, where work on India has mostly concentrated, to the incidence of compliance cost across firms, which can be followed in the public record; and it names the utilisation trap in shared infrastructure using the regulator's own figures. What follows is a mechanistic account rather than a causal-identification study, since enforcement is not exogenous and no clean experiment is available, and it is offered as a plausible and partly documented process rather than a measured effect. Tirupur serves as the empirical backbone and Jajmau as the contemporary test. Section~\ref{sec:background} sets out the regulatory and cost context, Section~\ref{sec:theory} the theory of change, Section~\ref{sec:method} the method, Section~\ref{sec:results} the results, Section~\ref{sec:mechanism} the mechanism, Section~\ref{sec:implications} the broader implications and boundaries, Section~\ref{sec:policy} the policy recommendations, and Section~\ref{sec:conclusion} concludes.

\section{Background}
\label{sec:background}

\subsection{How ZLD came to bind}
ZLD did not enter India's wet-processing clusters through a single statute that a legislature debated and passed. The obligation accumulated instead through pollution-control board directions and, in the textile clusters above all, through court orders, with the Madras High Court fixing the technical form for Tirupur in the mid-2000s as reverse osmosis followed by reject management and evaporation \citep{gronwall2017impact}. The two clusters examined in this paper sit on either side of the same industrial category. Tirupur is a textile centre built on dyeing and bleaching, while Jajmau, in Kanpur, is a leather cluster built on tanning. Both process wet, both release saline and chemically loaded effluent, and both are dominated by small units held to the same standard, which is what allows them to be read together. The manner of arrival matters for the argument that follows: because the obligation grew out of litigation about a river, no one examined how its cost would fall across the firms required to meet it.

Enforcement then runs through a layered set of institutions rather than a single regulator. The Central Pollution Control Board and its state counterparts monitor effluent and issue directions, including closure, under the Water Act of 1974 and the Environment (Protection) Act of 1986. In the Ganga basin, the National Green Tribunal and the higher courts have increasingly set the pace at which those powers are exercised. The division is worth holding in view, since the body that imposes a standard, the body that measures compliance with it, and the body that orders a plant or a firm to shut are not the same, and none of them is equipped to weigh the standard's incidence.

\subsection{Why compliance runs through shared plants}

The expense of a ZLD system is not merely large but poorly divisible. Reverse osmosis recovers most of the water, yet the evaporators and crystallisers that capture the remaining concentrate roughly double the cost of the plant and are found mainly at larger units running their own facilities \citep{gronwall2017impact}. A small firm cannot buy a proportionally smaller version of this train, because its minimum workable configuration is large relative to what one small unit produces. The response written into policy was the common effluent treatment plant, shared among clusters of units that, in the board's own account, cannot carry the capital and maintenance costs alone \citep{cpcb2014tirupur}. Adoption followed that logic. In Tirupur, somewhere between 80 and 110 units ran individual plants while roughly 350 were distributed across 18 shared ones, and the units with their own plants were predominantly the export-oriented firms whose foreign buyers pressed higher standards \citep{gronwall2017impact}.

\begin{figure}[H]
    \centering
    \includegraphics[width=0.85\linewidth]{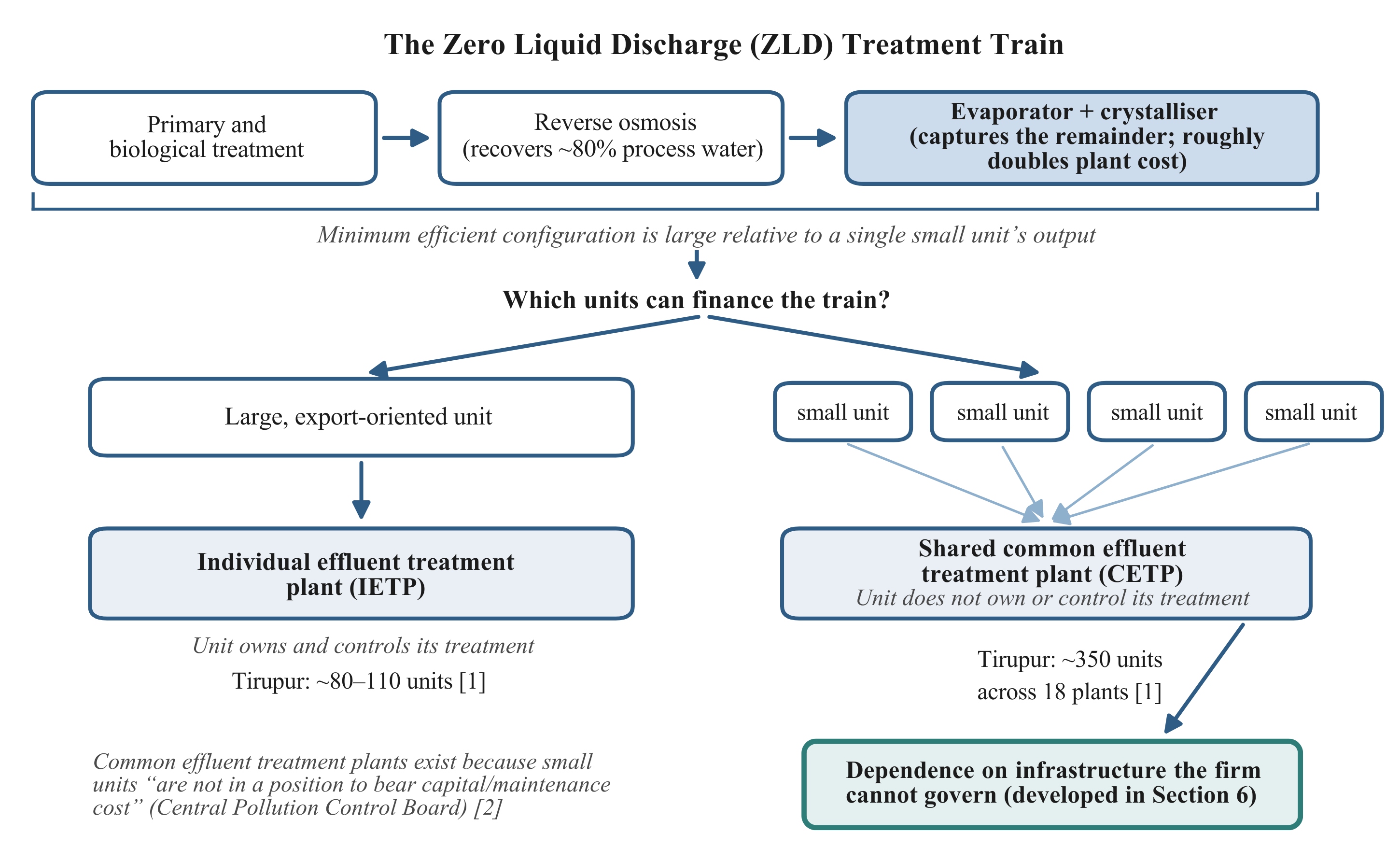}
    \caption{\textit{Two routes to ZLD, and the sorting they produce.} The treatment train runs left to right, with the evaporator-and-crystalliser back-end (shaded) roughly doubling plant cost; because that back-end cannot be scaled down, its minimum efficient size is large relative to a single small unit's output. Larger, export-oriented units finance individual plants (IETPs), while small units are sorted onto a shared common effluent treatment plant (CETP) they neither own nor control, leaving them dependent on infrastructure whose failure is developed in Section~\ref{sec:mechanism}. Indicative Tirupur counts ($\sim$80--110 units on IETPs; $\sim$350 across 18 CETPs) from \citep{gronwall2017impact}; the institutional rationale, that shared plants exist because small units cannot bear the capital and maintenance cost, from \citep{cpcb2014tirupur}.}
    \label{fig:routes}
\end{figure}

Shared plants carry a fragility that individual ones do not. Their running costs have to be recovered from many members at once, and the documented history of the model in India is one of recurring operation-and-maintenance and cost-recovery difficulty \citep{cpcb2006performance,kathuria2014small}. Why that difficulty tends to compound is a matter for Section~\ref{sec:mechanism}; here it is enough to record that dependence on a shared plant is dependence on an arrangement with a known tendency to underperform. The strength of the underlying cost claim should also be stated plainly. No measured curve of compliance cost against plant capacity appears to exist for India, so indivisibility rests on the regulator's stated rationale, on the sorting just described, and on a single abatement-cost study from a different industrial estate that found scale economies and a high marginal cost of treatment for small factories \citep{goldar2001water}. Whatever the precise shape of that curve, the consequence is not in doubt: a small unit's compliance comes to depend on infrastructure it neither owns nor governs.

\subsection{What the literature explains, and what it leaves open}

Three bodies of work bear on the question, and each is strong within its own boundaries. Research on abatement economics establishes the scale economies and the heavy burden borne by small industry \citep{goldar2001water}, while assessments of the shared-plant model find its performance persistently unsatisfactory, with only a small fraction of India's common effluent treatment plants meeting their standards even after two decades of operation \citep{cpcb2006performance,kathuria2014small}. On enforcement, empirical study shows that regulatory discretion is wide and consequential, and at times better aimed at serious polluters than a random inspector would be \citep{duflo2018value}, though monitoring tends to rest more lightly on the more profitable firm \citep{gupta2019environmental}. On measurement, the distinction between regulating equipment and regulating outcomes has a long theoretical pedigree \citep{besanko1987performance}, and field evidence from India shows that a compliance record can be manufactured when the party doing the measuring has reason to misreport \citep{duflo2013truthtelling}.

Yet these strands rarely connect. This literature treats cost, enforcement and measurement as separate determinants of who complies, not as consecutive stages of one process that begins with a firm unable to treat its own effluent and ends with that same firm entered in the record as a violator. A second gap concerns the object of study. The difficulty small industry faces under environmental enforcement is itself long recognised \citep{dasgupta2000environmental}, though it has been framed chiefly as a problem of poverty and regulatory capacity rather than as a question of how a shared compliance system distributes exposure across firms. Work on environmental justice in India, meanwhile, has concentrated on the communities exposed to pollution and on the workers displaced when polluting units are closed \citep{baviskar2006rethinking,bhuwania2018case}, and comparatively little of it follows the incidence of compliance cost across firms, even though that incidence is the more legible quantity in the public record of directions, audits and orders. Even the closest prior study of Tirupur's small bleaching and dyeing firms framed the question as whether collective action could improve their environmental performance, rather than as how the cost of compliance is distributed when the shared plant fails \citep{crow2006clean}. The background assembled here is meant to support precisely that composition and nothing further, since the mechanism developed in Section~\ref{sec:mechanism} turns on these three forces operating in sequence rather than in isolation.

\section{Theory of Change}
\label{sec:theory}

\subsection{Why this problem, and why now}

Indian environmental governance offers no shortage of problems to study, from urban air quality to municipal waste. This paper takes up a narrower one because it is at once consequential, live, and neglected. It is consequential because the decision at stake, whether to close a firm, is irreversible for the livelihoods involved and, on the argument advanced here, may not even deliver the cleaner river the closure was meant to secure. It is live because these decisions are being taken now, in the National Green Tribunal's continuing supervision of Ganga pollution and in the directions the pollution boards issue under it. And it is neglected because the incidence of compliance cost, the question of who ends up bearing a shared standard, has drawn far less attention than either the engineering of treatment or the exposure of nearby communities. A problem that is high-stakes, active, and under-examined is where a small, well-aimed study can do the most.

\subsection{Why it is tractable}

The problem is also more tractable than most in this field, which is a large part of why it is worth trying to solve rather than only to describe. The reforms the mechanism points towards are informational and procedural, not capital-intensive: they turn on what compliance is measured against, and in what order enforcement proceeds, rather than on any new treatment technology. The instruments they would use largely exist already, from the continuous effluent monitoring that polluting industries are required to run to the performance-linked operating contracts now used for sewage in the same river basin. And the decisions run through a small, identifiable set of institutions rather than a diffuse market, so influence does not mean moving millions of actors, only a few well-placed ones. A cheap fix, existing tools, and a short list of decision-makers put the problem within reach of a study like this.

\subsection{Who would act, and how the work could reach them}

The institutions that would have to act are readily named. The Central Pollution Control Board and the state boards compile the compliance record and issue directions; the National Green Tribunal and the higher courts set the pace of enforcement across the Ganga basin; the National Mission for Clean Ganga finances and contracts treatment infrastructure; effluent-plant operators and industry associations run and depend on the shared plants; and civil-society researchers, the Centre for Science and Environment and outlets such as Down To Earth among them, carry technical evidence into the public debate these bodies follow.

The route from research to influence is correspondingly concrete, and it turns on a single feature of the work: the argument is built almost entirely from the regulators' own performance records, audits and court orders. That provenance is what makes it awkward to dismiss and easy to route back to its source. A short brief drawn from the analysis can enter the same channel that already shapes Ganga enforcement, namely the tribunal's proceedings, which regularly take in expert and intervener submissions, and the civil-society reporting that amplifies them. The ask that follows is deliberately modest and specific: that a plant's performance be established before its member units are closed, and that the discharge data already collected, rather than the ownership of equipment, decide compliance. A change of that kind needs no new legislation, only a different use of powers the boards and the tribunal already hold, which is the sort of change they make on the strength of their own evidence.

\subsection{Assumptions, and the honest limits of the pathway}

We assume that decision-makers are responsive to evidence assembled from their own record; that a brief can reach the tribunal-and-civil-society debate rather than sit unread; and that the mechanism traced for Tirupur carries to the live case in Kanpur. None of these is guaranteed, and reform here is slow and shaped by many hands, so the claim is to a plausible pathway of influence, not a certain result. It is credible because it is modest: a specific reform, addressed to a few identifiable actors, argued from evidence they cannot easily reject, at a decision point that is open now. 

\section{Methodology}
\label{sec:method}

\subsection{Design and case selection}

The paper sets out to establish how a rule produces an outcome. Therefore, the method is a comparative case study built around process tracing \citep{george2005case,beach2013process}. Process tracing identifies the intervening steps that connect a proposed cause to its effect, then checks whether evidence shows each step actually occurred \citep{collier2011understanding}. It suits the present question for two reasons. The claim is mechanistic and focuses on the process by which a firm’s failure to treat its own effluent leads to its appearance in the compliance record. This process is confirmed or rejected by evidence from the case itself, rather than by comparing averages. An enforcement shock with an unusually complete documentary trail is also available, whereas a clean quasi-experiment, with exogenous variation in who is required to comply, is not.

Other designs were possible and were set aside for specific reasons. An econometric study would need exogenous variation in who must comply and firm-level panel data on costs and closures; neither exists for these clusters, and because enforcement is endogenous, a regression would recover selection as readily as the effect. A legal analysis would capture the orders but not the economics that drive them. An interview-based or ethnographic study, valuable for lived detail, could not establish the system-level incidence the question turns on, and would carry its own problems of access and recall. Process tracing fits a mechanistic question to the evidence actually available, a documented shock and a public paper trail, and the study is desk-based for the same reason. The decisive material is already public, which makes the argument feasible and, as Section~\ref{sec:theory} argues, harder for the institutions that produced it to wave away.

Two clusters carry the analysis in different roles. Tirupur, the Tamil Nadu textile centre, is the primary retrospective case. Its enforcement shock has already played out, and its records are detailed enough to trace the process from beginning to end. These include a peer-reviewed study, plant-level pollution-board data, and court orders. Jajmau, the Kanpur tannery cluster, is the contemporary prospective case. It faces a similar combination of shared treatment, poor performance, and stricter enforcement, but in a different sector. This allows the study to test whether the mechanism applies elsewhere while using the theory of change in Section~\ref{sec:theory} to examine an enforcement decision that is still open. Pairing a case where the mechanism has operated with one where it is operating guards against mistaking the idiosyncrasies of a single instance for the mechanism itself.

The design has a firm limit, stated here rather than deferred. Enforcement is not exogenous, so nothing in what follows estimates a treatment effect. The aim is to establish a plausible and documented mechanism and to specify, in advance, the observations that would overturn it.

\subsection{Data}

The evidence is drawn almost entirely from the public record, and the choice is deliberate. An argument assembled from the regulators' own material is harder for the regulators to dismiss. The government sources are load-bearing. They comprise the Central Pollution Control Board's annual reports and its assessments of common effluent treatment plant performance \citep{cpcb2017annual,cpcb2006performance}; the Board's techno-economic report on Tirupur, which gives plant-by-plant capacity, throughput and treatment charges \citep{cpcb2014tirupur}; a Comptroller and Auditor General performance audit of the relevant environmental facilities in Uttar Pradesh \citep{cag2017report}; and the tribunal and court orders that imposed and enforced the standard \citep{gronwall2017impact,ngt2021jajmau}. Peer-reviewed work supplies the Tirupur transition narrative and the economics of abatement \citep{gronwall2017impact,goldar2001water}. Reporting by civil-society bodies is used sparingly, only where it fills a gap the official record leaves open, and it is flagged as journalism rather than treated as data. Where such reporting appears, as with the contemporaneous estimate that river salinity fell only partially after the Tirupur closures \citep{seth2011towards}, it illustrates a point the official record leaves undocumented rather than settling it.

No primary fieldwork, including interviews and firm-level surveys, was undertaken. Where a concrete detail of firm behaviour appears, it serves as an illustration and is labelled as such. One feature of the data deserves emphasis, because it is part of the object of study and not merely a constraint on it. Some of the key evidence comes from records created by faulty measurement systems, especially the defunct flow meters at Jajmau \citep{cag2017report}. Since the mechanism predicts that poor measurement itself can become part of the compliance problem, these records are treated as evidence.

\subsection{What would count as evidence, and what against}

A mechanism claim earns credibility by committing in advance to what each stage should leave behind if it is real, and to what observation would tell against it. Table~1 sets this out stage by stage. The indivisibility stage predicts that firms choose between individual and shared treatment based on their size and export orientation, with smaller firms relying more on shared treatment. It would be weakened if ZLD costs stayed proportional across different plant sizes. The utilisation stage predicts that treatment costs per unit rise as a plant becomes less utilised, and that closures increase costs for the firms that remain. It would be weakened if costs did not change with utilisation. The sorting stage predicts that firms with in-house treatment are more likely to survive an enforcement shock. It would be challenged if survival instead depended on political connections regardless of firm size or treatment capacity. The measurement stage, the best supported of the four, predicts that compliance is scored on the possession of equipment and that recorded status parts company with measured discharge. Some of these are hoop tests, which a stage must pass to stay credible, while others come closer to smoking-gun tests, whose presence would strongly confirm a stage even where their absence would not sink it \citep{collier2011understanding}; the columns of Table~1 are arranged to keep the two apart.

\begin{table}[H]
    \caption{\textit{The mechanism's observable implications and defeaters.} For each stage of the mechanism: what should be observed if it holds, the evidence relied on, and the single observation that would count against it. The columns separate hoop tests, which a stage must pass to remain credible, from smoking-gun tests, whose presence would strongly confirm a stage.}
    \label{tab:predictions}
    \centering
    \small
    \rowcolors{2}{gray!12}{white}
    \begin{tabularx}{\textwidth}{@{}>{\raggedright\arraybackslash}p{0.16\textwidth} >{\raggedright\arraybackslash}X >{\raggedright\arraybackslash}X >{\raggedright\arraybackslash}p{0.19\textwidth}@{}}
        \toprule
        \textbf{Mechanism stage} & \textbf{Observable implication if true} & \textbf{Evidence used} & \textbf{What would count against it} \\
        \midrule
        1. Cost indivisibility strands small firms
            & Firms sort between individual and shared treatment by size and export orientation; no proportionally small ZLD plants appear
            & Tirupur sorting of units between IETPs and CETPs \citep{gronwall2017impact}; regulator's stated rationale \citep{cpcb2014tirupur}; abatement scale economies \citep{goldar2001water}
            & A measured ZLD cost curve flat across plant capacity \\
        \addlinespace
        2. Underuse degrades (utilisation trap)
            & Treatment cost per unit rises as utilisation falls; closures feed back into lower utilisation and higher charges on remaining members
            & CPCB plant-level charge and utilisation figures, Tirupur \citep{cpcb2014tirupur}; record of O\&M and cost-recovery failure \citep{cpcb2006performance,kathuria2014small}
            & Cost per unit invariant to utilisation, or closures not followed by higher member charges \\
        \addlinespace
        3. Absorptive capacity sorts who survives
            & Survivors of the shock concentrated among firms with in-house treatment, typically larger and export-oriented
            & Composition of units permitted to continue vs closed \citep{gronwall2017impact,cag2017report,ngt2021jajmau}; enforcement heavier on some firm types \citep{duflo2018value,gupta2019environmental}
            & Survival tracks political connection independently of firm size and treatment capacity \\
        \addlinespace
        4. Measurement misattributes failure
            & Compliance recorded as possession of equipment; recorded status diverges from measured discharge
            & Design vs performance standard theory \citep{besanko1987performance}; audit misreporting evidence \citep{duflo2013truthtelling}; defunct metering at Jajmau \citep{cag2017report}
            & Compliance consistently scored on measured discharge, tracking actual effluent quality \\
        \bottomrule
    \end{tabularx}
\end{table}

These tests are not equally decisive, and Table~1 records the differing weight of each stage, and its potential defeater, rather than leaving them implicit.

\subsection{Limitations and threats to inference}

The desk-based design brings clear limits. Without firm-level data, incidence is read from an aggregate public record that is coarse and, for the tannery case, incomplete. As enforcement is endogenous, no causal identification is claimed. The interpretive limits that follow, including the risk that the size gradient is partly an artefact of the measurement the fourth stage criticises, are taken up in Section~\ref{subsec:limitations}. The findings that follow are offered as evidence for a plausible and partly documented mechanism, not a measured effect.

\section{Results}
\label{sec:results}

Firms sort by size into different treatment arrangements. In Tirupur, somewhere between 80 and 110 units ran their own effluent treatment plants, while roughly 350 depended on 18 shared ones. The units with individual plants were predominantly the larger, export-oriented firms \citep{gronwall2017impact}.

\begin{figure}[H]
    \centering
    \includegraphics[width=0.85\linewidth]{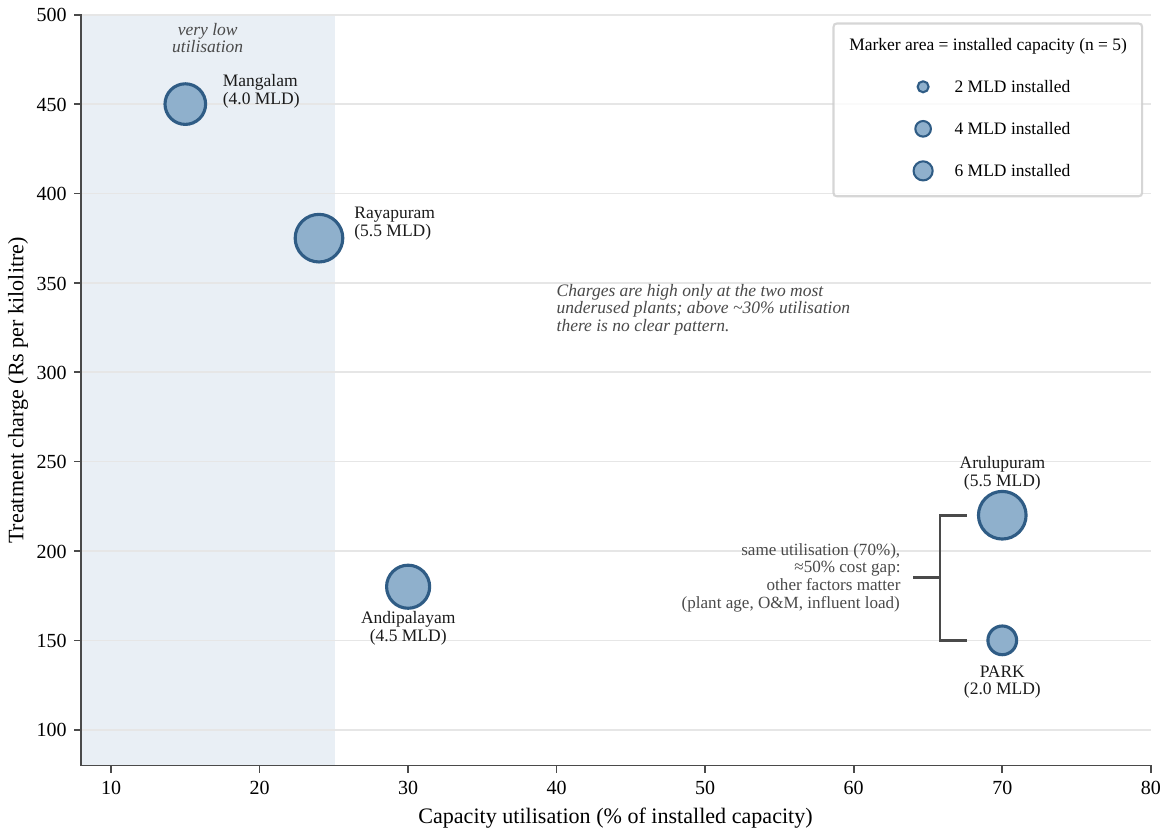}
    \caption{\textit{Per-unit treatment cost against capacity utilisation across five Tirupur CETPs (n = 5)}. Marker area is proportional to installed capacity (MLD). High charges appear only at the two most underused plants; above about 30 per cent utilisation, the cross-section is flat, and two plants at the same 70 per cent utilisation differ by roughly half, indicating factors beyond utilisation such as plant age and influent load. Source: CPCB 2014--15 \citep{cpcb2014tirupur}.}
    \label{fig:cost}
\end{figure}

Per-unit cost spikes at the most underused plants. In the pollution board's plant-level figures, the two most underused plants, at 15 and 24 per cent, charged Rs 450 and Rs 375 per kilolitre, while the three running at 30 per cent or above charged between Rs 150 and Rs 220 (Figure~\ref{fig:cost}). The relationship is not linear. Above roughly 30 per cent utilisation the charges are flat and noisy, and two plants at the same 70 per cent differ by about half, pointing to factors beyond utilisation such as plant age and influent load. Installed capacity does not order the costs either, since the cheapest plant is among the smallest and the largest appears at both ends \citep{cpcb2014tirupur}. With five plants, this is a descriptive pattern, not a measured relationship.

The enforcement shock fell unevenly by size. When the 2011 closures came, units with in-house treatment were placed to resume, while many smaller units tied to shared plants were not. About 700 units shut down, close to 100 never reopened, and some production returned at micro-scale in residential buildings discharging into the municipal sewer \citep{gronwall2017impact}.

The same configuration is failing now in Kanpur. The Jajmau tannery plant has been recorded as non-complying across successive pollution board annual reports \citep{cpcb2017annual}. A state audit found it treating a fraction of the effluent reaching it, releasing water that failed irrigation standards, and running with defunct flow meters \citep{cag2017report}.

Closure did not visibly clean the water. After the Tirupur shutdowns, one contemporary estimate put the fall in river salinity at only partial, from about 5,000 to 3,000 mg per litre against a norm near 250. The figure is journalistic and confounded by legacy and upstream load, and is reported with that caution \citep{seth2011towards}.

Read together, the findings describe a burden that settles on small, shared-plant-dependent firms, tied to infrastructure whose cost depends on how fully it is used and whose failure is entered against its members. What none of these observations yet explains is why the burden falls along exactly this line, and why closing the firms relieves neither them nor the river. That is the question Section~\ref{sec:mechanism} takes up.

\section{Discussion: constructing the mechanism}
\label{sec:mechanism}

\subsection{One process, not three explanations}

The literature offers three candidate accounts (cost, politics, and measurement), and each advanced on its own. The argument here says that they are stages at different positions in one process that runs from a firm's inability to treat its own effluent to its entry in the record as a violator. Taken separately, each explains a fragment. Only in sequence do they explain the shape of the whole. The individual layers are largely already present in the literature. Their composition into a single sequence is not, nor is one link within it. The layers also differ in how firmly they are grounded, and each is marked below by the kind of claim it rests on: demonstrated, inferred, assumed, or conjectured.

\subsection{The generative layer: cost indivisibility}

Fixed-cost indivisibility is the layer that generates the others, in the precise sense that removing it dissolves them. Were a small unit able to treat its own effluent at proportional cost, it would not depend on a shared plant. It would a;so not be exposed to that plant's failure, and would never be recorded as non-compliant on its account. Dependence clearly follows from the shape of the cost, not from any choice the firm makes. The sorting observed in Section~\ref{sec:results}, with larger and export-oriented firms using individual plants and smaller firms using shared plants, matches the indivisibility prediction and is supported by the record \citep{gronwall2017impact,cpcb2014tirupur}.

The claim is inferred as no measured cost curve against plant capacity exists for India. The primacy assigned to cost rests on the regulator's own rationale for shared plants, on the observed sorting, and on a single abatement-cost study from a different setting \citep{cpcb2014tirupur,goldar2001water}. Indivisibility at small scale is thus an assumption the available evidence supports but does not measure.

\subsection{The failure layer: the utilisation trap}

A revealing feature of the pattern is that the two most underused plants are the dearest per unit, while installed capacity does not order the costs. Economies of scale would have cost falling as capacity grows. Instead, the cheapest plant is among the smallest, and above roughly a third of capacity the cross-section is flat. A shared plant spreads its fixed costs across throughput, so a plant running far below capacity becomes markedly dearer for each member who stays.

From this follows a self-reinforcing failure that a static reading would miss. As charges climb on a draining plant, marginal members default or withdraw, throughput falls again, and the charge rises once more, until the plant drops below the load it needs to hold its own standards. The causation plausibly runs in both directions, since a costly and badly run plant also sheds members, and their exit raises the charge again. The cross-section cannot separate the two, but both drive the same loop. The loop also explains a detail the Tirupur record otherwise leaves unexplained: operators who had built better treatment reportedly slowed to the pace of weaker ones, because the extra investment left them worse off while charges were set in common \citep{gronwall2017impact}. It also implies that removing members from a struggling shared plant lowers utilisation for those who remain, creating a feedback loop in which enforcement against individual units can contribute to the failure of the shared plant. The implications of this feedback for closure decisions are examined in Section~\ref{sec:policy}; here, the point is that the evidence links the different stages of the mechanism rather than treating them as independent processes.

The trap is established descriptively, from a handful of plants in one cluster at a single moment, and stands as a suggestive cross-section rather than a demonstrated dynamic \citep{cpcb2014tirupur}. It is the novel step in the mechanism, which Figure~\ref{fig:mechanism} marks as a contribution.

\subsection{The allocation layer: absorptive capacity, and why not politics}

When enforcement tightened, the firms that carried on were those with in-house treatment, generally larger and export-oriented \citep{gronwall2017impact}. The most economical reading is capacity to absorb a shock. A firm that already controls its own compliance is not hostage to a shared plant, and the export orientation of the survivors is itself telling, since exporting firms abate more where they can spread the fixed costs of abatement across greater output \citep{forslid2018why}. That is why the burden is graded by size, and it issues from the same cost structure as the earlier layers. 

A political account would trace survival instead to connection and influence. It cannot be ruled out and may operate at the margin, yet three considerations make it the weaker primary explanation. It does not predict why vulnerability concentrates in precisely the firms that depend on shared treatment, which is the fact most in need of explaining. The strongest Indian evidence on regulatory discretion suggests that enforcement is more effectively targeted at serious polluters than random inspection would be, rather than being systematically captured \citep{duflo2018value}. A large study of firm characteristics also finds that enforcement falls more lightly on more profitable firms, suggesting that profitability, rather than size or political connections, may shape enforcement \citep{gupta2019environmental}. Evidence that regulatory burden itself falls on smaller firms, whose marginal entrant is markedly smaller where permit conditions bind, tends in the same direction \citep{kala2025environmental}. Political selection is therefore retained as a possible amplifier, thinly evidenced, and not leaned upon, as Figure~\ref{fig:mechanism} records.

\subsection{The epistemic layer: measurement}

The last layer explains why the outcome is entered as the small firm's own non-compliance. Where compliance is scored on the possession of working treatment rather than on measured discharge, a firm stranded by a failing shared plant appears in the record as an offender in its own right. This is best-supported as it does not rest on the present cases alone. The distinction between design and performance standards is long established in theory \citep{besanko1987performance}, and experimental evidence from India shows that compliance records can be manufactured when the party doing the measuring has reason to misreport \citep{duflo2013truthtelling}. The defunct flow meters at Jajmau are the same failure in physical form, a plant sitting in judgement on its members while unable to measure what passes through it \citep{cag2017report}. That measurement is an active determinant, not a passive record, shows most clearly where an Indian emissions market replaced lax monitoring with continuous metering and compliance rose from roughly two-thirds of plants to nearly all \citep{greenstone2025pollution}. Measurement neither creates the dependence nor causes the plant to fail. It conceals both, and in concealing them it makes the resulting distribution look deserved.

\subsection{Why the composed account fits better than its parts}

As opposed to its alternatives, the layered mechanism explains a slice of the pattern while the sequence (layers) explains its shape. Table~2 sets the candidate accounts against the observations. A cost-only story handles the sorting but neither the utilisation relationship nor the misattribution. A politics-only story can rationalise selective survival yet not the concentration of exposure in shared-plant-dependent firms. A measurement-only story explains the misattribution but not the exposure it misattributes. The layered process reaches all of them.

\begin{table}[H]
    \caption{\textit{Which explanation accounts for which observed pattern.} Rows are the five patterns established in Section 5; columns are the three single-factor accounts and the composed mechanism. A filled circle marks a pattern an account explains on its own, a half circle a partial or conditional fit, and an open circle a failure to account for it. Only the composed mechanism reaches every pattern.}
    \label{tab:accounts}
    \centering
    \small
    \rowcolors{2}{gray!12}{white}
    \begin{tabularx}{\textwidth}{@{}>{\raggedright\arraybackslash}X
        *{4}{>{\centering\arraybackslash}p{0.135\textwidth}}@{}}
        \toprule
        \textbf{Observed pattern (Section 5)} & \textbf{Cost-only (scale economies)} & \textbf{Politics-only (capture)} & \textbf{Measurement-only (data artefact)} & \textbf{Composed mechanism} \\
        \midrule
        Firms sort by size into individual vs shared (CETP) treatment & \ballfull & \ballempty & \ballempty & \ballfull \\
        Treatment cost tracks utilisation, not plant size & \ballhalf & \ballempty & \ballempty & \ballfull \\
        Survival after the shock is size-graded & \ballhalf & \ballhalf & \ballempty & \ballfull \\
        Failure recorded as the small firm's own non-compliance & \ballempty & \ballhalf & \ballfull & \ballfull \\
        Closure does not visibly clean the river & \ballempty & \ballempty & \ballhalf & \ballhalf \\
        \bottomrule
    \end{tabularx}

    \vspace{2pt}
    {\footnotesize \ballfull~accounts for it \quad \ballhalf~partial / conditional \quad \ballempty~does not account for it}
\end{table}

The account stays defeasible. A measured cost curve flat across capacity would remove the generative layer, and the ordering would fall apart. Survival tracking political connection independently of size and treatment capacity would move allocation ahead of cost. The weakest observation, that closure did not visibly clean the river, is confounded by legacy and upstream load and can bear no more than illustrative weight for now \citep{seth2011towards}, though it is at least consistent with national evidence that India's water-pollution regulations have produced no measurable benefit, in contrast with its air rules \citep{greenstone2014environmental}. What is offered is not a proven chain but a composition that fits the evidence better than any single-factor rival, with its weak points named.

\subsection{What the mechanism implies for intervention}

Read this way, the familiar picture of small firms as an industry's dirty laggards nearly inverts. The units most likely to close are not the worst polluters but the ones a policy design placed downstream of a shared plant they cannot govern. Enforcement in these clusters runs in a settled order, assembling a compliance record largely from the possession of equipment and then closing the units found wanting. The mechanism implies that this order is close to reversed. Acting on the record first means relying on a measure that cannot distinguish between a firm that is polluting and one that is affected by a failed shared plant. Closure also reduces the plant’s utilisation, which can deepen the failure for the firms that remain. The first question, therefore, must be whether the plant or the firm has actually failed, based on measured discharge (Table~3). One objection is that verifying performance first could delay enforcement. The answer is not to weaken enforcement, but to assign responsibility correctly: a failing shared plant remains the responsibility of both its operator and the authority responsible for its operation.

\begin{table}[H]
    \caption{\textit{The mechanism reorders intervention.} Each row gives a layer of the mechanism in the priority order the argument implies, the question it answers first, why acting out of that order backfires, and the confidence behind it. Current practice inverts the order, acting on closure first and on the equipment-based record.}
    \label{tab:order}
    \centering
    \small
    \rowcolors{2}{gray!12}{white}
    \begin{tabularx}{\textwidth}{@{}>{\raggedright\arraybackslash}p{0.17\textwidth}
        >{\raggedright\arraybackslash}X >{\raggedright\arraybackslash}X
        >{\raggedright\arraybackslash}p{0.15\textwidth}@{}}
        \toprule
        \textbf{Priority order (mechanism-implied)} & \textbf{Question it answers first} & \textbf{Why acting out of this order backfires} & \textbf{Confidence} \\
        \midrule
        1. Measurement (epistemic layer)
            & Is compliance scored on measured discharge, or merely on owning treatment equipment?
            & Closures rest on a record that cannot tell a firm that pollutes from a firm stranded by a broken shared plant
            & Strong \citep{besanko1987performance,duflo2013truthtelling,cag2017report} \\
        \addlinespace
        2. Cost and shared-plant performance (generative + failure)
            & Has the shared plant or the firm actually failed, and can the plant be made to perform?
            & Removing members lowers utilisation and, via the utilisation trap, deepens the plant's failure for those who remain
            & Suggestive \citep{gronwall2017impact,cpcb2014tirupur} \\
        \addlinespace
        3. Procedural safeguards (allocation layer)
            & Is the incidence of enforcement fair once the system is measured and functioning?
            & Fair process applied over a mismeasured and still-failing system changes little on its own
            & Thin/contextual \citep{duflo2018value,gupta2019environmental} \\
        \bottomrule
    \end{tabularx}
\end{table}

\subsection{A live test: Kanpur}

Kanpur turns the argument from a retrospective reading into a live test. The Jajmau plant is recorded as non-complying across successive board reports and was found by a state audit to treat only a fraction of its inflow with instruments that did not work \citep{cpcb2017annual,cag2017report}, while dependent units sit under active tribunal and board pressure to close \citep{ngt2021jajmau}. The mechanism predicts that closing these firms will reduce utilisation of the shared plant they depend on, increasing the burden on the firms that remain. Yet the compliance record may continue to attribute the resulting problems to the firms themselves. The decision is therefore being made using exactly the measure that the mechanism identifies as uninformative.

\subsection{What would move the account from plausible to tested}

Each weak point in the mechanism implies a specific empirical test. A measured ZLD cost curve across plant capacities would provide the strongest test of the generative layer and represents the most valuable missing datum. A panel tracking utilisation and treatment charges over time would allow the trap to be tested dynamically rather than only through cross-sectional evidence. Firm-level survival data linking closure outcomes to firm size, treatment type, and political connections would help distinguish absorptive capacity from political selection, which the present analysis cannot fully resolve.

\section{Broader implications and boundaries}
\label{sec:implications}

\subsection{Where the mechanism might apply}

Set against the wider field, the contribution of the paper is not really about Tirupur or Jajmau. It is about a configuration. Wherever compliance is assessed on the individual firm but depends, in practice, on shared infrastructure the firm cannot govern, the possibility described here is present: a collective failure recorded as a private one. The common effluent treatment model is not peculiar to these two clusters. It is the standard Indian device for pollution control among small industry, adopted across many industrial estates precisely because small units cannot treat their own effluent \citep{cpcb2006performance}. The nearest and safest extension of the argument is therefore to other effluent clusters built on shared plants, where the same cost structure and the same measurement habit are likely to recur. Beyond this, the mechanism is best understood as specifying what evidence to look for, rather than asserting where that evidence will be found. Collective infrastructure, individual assessment and equipment-based measurement can co-occur in settings far from textiles and leather, but whether the failure follows there is an open empirical question that two cases cannot settle.

\subsection{The conditions under which it holds}

The mechanism should hold only where four conditions occur together. Compliance must require treatment that is genuinely indivisible at small scale, so that self-provision is uneconomic. That treatment must be shared and outside the individual firm's control. Its costs must be recovered from throughput, so that utilisation governs the price each member pays. And compliance must be scored on the possession of equipment rather than on measured discharge. Removing even one would loosen the account. A cluster whose firms can treat their own effluent cheaply, a shared plant that is well funded and professionally run, or a regime that measures what actually leaves the pipe would each break the chain at a different link. External validity is thus conditional rather than general, and the two cases here demonstrate the configuration operating twice.

\subsection{Limitations}
\label{subsec:limitations}

The limitations bear on interpretation in specific ways. Because the paper reads incidence from an aggregate public record, the size-graded burden is an inference from that record rather than an observation of firm-level outcomes. This matters because the record is partly the very thing under criticism here. If compliance is mismeasured, then the size gradient the argument treats as an outcome could itself be shaped by the measurement, and the two cannot be fully separated with the evidence to hand. That circularity is the deepest limitation, and it caps how strongly the allocation layer can be pressed. The utilisation trap carries a limitation of a different kind: it rests on a small cross-section from one cluster at one moment, which cannot distinguish the proposed dynamic from confounds such as plant vintage or influent load. None of this dissolves the contribution, which lies in bringing a scattered literature together into a single testable account. However, it means that the claim should be treated as plausible and partially supported rather than established. A further caveat concerns the framing of ZLD regulation. The argument treats its adoption as the result of regulatory direction and litigation rather than a single national statute, and this interpretation should be verified before drawing stronger conclusions.

\subsection{The research that would test it}

The most useful research from here looks outward. Section~\ref{sec:mechanism} named the tests that would probe the mechanism's internal joints; the questions that remain are about its reach. The first is replication across common effluent treatment plants that differ in utilisation and in how compliance is measured, which would show whether the configuration reliably produces the outcome or whether these two clusters are unrepresentative. The second is comparative by design. Setting a regime that scores compliance on discharge against one that scores it on equipment, allowing the independent contribution of the measurement layer to be isolated rather than inferred. The third asks whether the mechanism travels at all beyond effluent, to other domains where compliance leans on collective infrastructure, which would test the generality so far offered only as conjecture. Each of these targets external validity, and together they would settle whether this is a general account of shared-infrastructure compliance or an accurate description of two Indian clusters. The aim is to specify a mechanism that can be empirically tested while remaining clear about the limits of the evidence. Its broader validity depends on whether the mechanism recurs in other settings and whether its predictions are supported by further evidence.

\section{Policy Recommendations}
\label{sec:policy}

\subsection{Immediate: what compliance is scored on}

The immediate priority is to change how compliance is assessed. The strongest evidence from the mechanism suggests that treating the possession of equipment as proof of compliance can misattribute a shared plant’s failure to its member firms. Compliance and closure decisions should therefore be based on measured discharge rather than installed treatment capacity. The instrument for this already exists. Since 2014 the Central Pollution Control Board has required tanneries, textile units and common effluent treatment plants to install online continuous effluent monitoring systems (OCEMS) that transmit discharge data in real time \citep{cpcb2018ocems}. The reform is therefore not to build new machinery but to make that discharge record, rather than equipment ownership, the basis on which the boards and the National Green Tribunal decide who is complying and who may be closed. Where a plant's own monitoring has failed, as at Jajmau, whose flow meters were found defunct \citep{cag2017report}, that failure should be read as the plant's, and metering restored, before any member unit is penalised.

\subsection{Medium-term: repairing the shared plant}

The medium-term steps address the plant itself. Because withdrawing members from an underused plant can deepen its failure for those who remain, closure of a member unit should be conditional on an independent finding that the fault lies with the firm rather than the shared plant. This responsibility falls to tribunals, courts, and state boards, and follows from the utilisation trap rather than from any general preference for leniency. The finding need not require a new system, since a performance audit like the one that exposed the Jajmau plant’s shortfall \citep{cag2017report} provides a natural basis for assessment. Such an audit should be conducted before a closure order rather than after it. Shared plants should also be placed on a performance-linked operating framework. The Hybrid Annuity Model adopted for sewage treatment under the Namami Gange programme, which ties an operator's payment to measured performance over 15 years and consolidates fragmented plants under a single accountable operator, is already being applied to Kanpur's municipal sewage \citep{ifc2023breathing}. Extending that structure to tannery and textile effluent plants would attack the operation-and-maintenance failure the mechanism locates at the plant. However, this model is proven for sewage, not yet for industrial effluent. Currently, it is offered as a transferable instrument rather than a demonstrated fix. A narrower measure, viability-gap support or a utilisation floor that stops one member's exit from spiralling charges onto the rest, targets the trap more directly but rests on the same descriptive cross-section, and should be piloted before it is mandated.

\subsection{Longer-term: capacity and procedure}

The longer-term steps concern the conditions needed to implement the earlier reforms. Measurement reform cannot succeed without the capacity of the boards responsible for enforcement. The documented pattern of staff vacancies and polluter-weighted membership makes institutional capacity and independence at state boards a precondition for effective reform \citep{ghosh2022state}. A further procedural safeguard would require reasoned orders to distinguish plant-caused from firm-caused non-compliance and give affected firms an opportunity to be heard before closure. This would formalise the reattribution required by the mechanism. The record of mass closures without such hearings illustrates the consequences of their absence \citep{bhuwania2018case}. This is the least well-supported recommendation and is therefore presented as a procedural implication of the argument rather than as a conclusion independently established by the data.

\begin{table}[H]
    \caption{\textit{Recommendations by sequence, stakeholder, evidence and obstacle.} Each recommendation follows from a specific layer of the mechanism and is placed in the order set out in Section 6. The evidence column carries the paper's calibration, so the strength behind each recommendation is visible rather than implied.}
    \label{tab:reforms}
    \centering
    \footnotesize
    \rowcolors{2}{gray!12}{white}
    \begin{tabularx}{\textwidth}{@{}>{\raggedright\arraybackslash}X
        >{\raggedright\arraybackslash}p{0.12\textwidth}
        >{\raggedright\arraybackslash}p{0.085\textwidth}
        >{\raggedright\arraybackslash}p{0.13\textwidth}
        >{\raggedright\arraybackslash}X@{}}
        \toprule
        \textbf{Recommendation} & \textbf{Directed at} & \textbf{Sequence} & \textbf{Evidence} & \textbf{Main implementation challenge} \\
        \midrule
        Decide compliance and closure on measured discharge from the already-mandated OCEMS, not on equipment
            & CPCB, state boards, NGT & Immediate
            & Strong \citep{duflo2013truthtelling,greenstone2025pollution,cpcb2018ocems}
            & Switching the decision basis; data quality and meter uptime \\
        \addlinespace
        Treat a plant's failed metering as the plant's non-compliance; restore it before any member is closed
            & State boards, NGT & Immediate
            & Strong \citep{cag2017report,cpcb2018ocems}
            & Attribution disputes; funding and maintaining meters \\
        \addlinespace
        Condition member-unit closure on a verified finding that the fault is the firm's, not the shared plant's
            & NGT, courts, state boards & Medium
            & Suggestive \citep{gronwall2017impact,cpcb2014tirupur}
            & Slows enforcement; needs independent plant assessment \\
        \addlinespace
        Put CETP operation on a performance-linked footing (extend HAM / one-operator from sewage to effluent)
            & NMCG, state govts, funders, operators & Medium
            & Transferable, unproven for effluent \citep{ifc2023breathing}
            & Untested on industrial effluent; contracting capacity \\
        \addlinespace
        Viability-gap support or a utilisation floor so one member's exit does not spiral charges onto the rest
            & State govts, industry bodies, funders & Medium
            & Weak; descriptive trap \citep{cpcb2014tirupur}
            & Fiscal cost and moral hazard; pilot before mandating \\
        \addlinespace
        Strengthen state-board capacity and independence as a precondition for measurement reform
            & MoEFCC, state govts & Longer
            & Documented need \citep{ghosh2022state}
            & Political will; staffing and recruitment \\
        \addlinespace
        Require a reasoned order separating plant-caused from firm-caused failure, with a hearing, before closure
            & Courts, boards & Longer
            & Thin \citep{bhuwania2018case}
            & Judicial workload; risk of procedural delay \\
        \bottomrule
    \end{tabularx}
\end{table}

\subsection{Two cautions}

Firstly, the argument does not imply that closure is ineffective, only that firms should not be penalised for failures they did not cause or could not prevent. Secondly, the mechanism does not guarantee that the proposed reforms will succeed. It supports their direction and sequencing, but implementation depends on institutional and political factors beyond the analysis. The evidence is strongest for the first. Compliance should be assessed using measured discharge from existing instruments rather than installed capacity.

\section{Conclusion}
\label{sec:conclusion}

The question was why, when ZLD enforcement tightens, the burden settles on small firms rather than on the units that pollute most. The answer offered here is structural rather than moral. A shared system fails, and the way compliance is financed, allocated, and measured makes the failure look like theirs. Framing the problem as the incidence of a design, rather than the conduct of a firm, is the paper's contribution, and the utilisation trap is the part of that design most easily overlooked and most readily repaired by observation. The remedies follow from that diagnosis, and begin where it does, with what compliance is measured on. For a circular economy that presents itself as both clean and fair, the caution is real. Closing the weakest participants in a failing collective arrangement can move pollution without lessening it, and can penalise firms for a shortfall they had no power to prevent. Whether circularity is just depends on what one chooses to measure.

\section*{Acknowledgements}

This idea began during the Non-Trivial Fellowship, and I am grateful to the Non-Trivial team for the opportunity to pursue it. I thank my facilitator, Ulyana Bachtizina, for her guidance. I also thank Vick Volovnyk, Edward Cheung, and Aidan Gao for their feedback, and Dr. Kaibalyapati Mishra for the arXiv endorsement.

\bibliographystyle{unsrtnat}
\bibliography{references}

\end{document}